# Model for processive nucleotide and repeat additions by the telomerase

Ping Xie

*Department of Physics, Renmin University of China, Beijing 100872, China*

## Abstract

A model is presented to describe the nucleotide and repeat addition processivity by the telomerase. In the model, the processive nucleotide addition is implemented on the basis of two requirements: One is that stem IV loop stimulates the chemical reaction of nucleotide incorporation, and the other one is the existence of an ssRNA-binding site adjacent to the polymerase site that has a high affinity for the unpaired base of the template. The unpairing of DNA:RNA hybrid after the incorporation of the nucleotide paired with the last base on the template, which is the prerequisite for repeat addition processivity, is caused by a force acting on the primer. The force is resulted from the unfolding of stem III pseudoknot that is induced by the swinging of stem IV loop towards the nucleotide-bound polymerase site. Based on the model, the dynamics of processive nucleotide and repeat additions by *Tetrahymena* telomerase are quantitatively studied, which give good explanations to the previous experimental results. Moreover, some predictions are presented. In particular, it is predicted that the repeat addition processivity is mainly determined by the difference between the free energy required to disrupt the DNA:RNA hybrid and that required to unfold the stem III pseudoknot, with the large difference corresponding to a low repeat addition processivity while the small one corresponding to a high repeat addition processivity.

*Keywords*: telomerase; nucleotide addition processivity; repeat addition processivity; molecular motor

# 1. Introduction

Telomerase is a specialized cellular ribonucleoprotein (RNP) complex that uses an intrinsic RNA to template the de novo synthesis of telomeres onto the 3' ends of linear chromosomes [1–5]. The telomerase activity was shown firstly for the ciliate *Tetrahymena thermophila* [6] and subsequently for many other eukaryotes [7–16]. Telomerase RNP functions as a multisubunit holoenzyme that contains an RNA component (telomerase RNA), a catalytic protein component telomerase reverse transcriptase (TERT) and other associated proteins [17]. *Tetrahymena* telomerase RNA is a 159 nucleotides transcript, in which contains the sequence 3'-AACCCCAAC-5' that serves as a template for synthesis of the telomeric repeat TTGGGG [6]. Immediately located 5' of the template is a template boundary element (TBE) and 3' of the template is a template recognition element (TRE) [18,19]. In addition to these template-adjacent elements, the pseudoknot III and stem-loop IV have also been shown to play important roles in telomerase function [18, 20–22]. The TERT protein contains polymerase site motifs shared among all reverse transcriptases [23] as well as unique N- and C-terminal extensions that harbor both phylogenetically conserved and variable motifs [24]. In addition, it has been shown that the TERT protein alone provide the anchor site for the telomerase that interacts with the primer [25–27]. Of the accessory proteins, p65 has been shown to form a complex with telomerase RNA and plays role in assembly of telomerase RNP [28–30].

A peculiar feature of the telomerase is its ability to synthesize long stretches of a primer by using the short template sequence within telomerase RNA [4,6,25,31]. This requires that the telomerase must be able to carry out two types of translocation reaction: the forward translocation of the polymerase site along template after each nucleotide incorporation (nucleotide addition processivity), and the disruption of the DNA:RNA hybrid and repositioning of the product 3' end to the beginning of the template after each round of repeat synthesis (repeat addition processivity). However, up to now, the detailed molecular mechanism of these two processive additions still remains unclear.

In this work, based on the available biochemical and structural studies of telomerase and on the analogy of previously proposed mechanisms for the processive nucleotide addition by other polymerase enzymes such as DNA polymerase (DNAP) [32] and RNA polymerase (RNAP) (Xie, unpublished), a model is presented for these processive nucleotide and repeat additions by the telomerase. In the model, during the processive nucleotide addition the forward movement of the polymerase site along the template is rectified through the incorporation of a matched base, in a Brownian ratchet mechanism. The unpairing of DNA:RNA hybrid after the incorporation of the nucleotide paired with the last base on the template, which is the prerequisite for repeat addition processivity, is caused by a force acting on the primer. The force is resulted from the unfolding of stem III pseudoknot that is induced by the swinging of stem IV loop towards the nucleotide-bound polymerase site. Thus the repeat addition processivity is mainly determined by the difference between the free energy required to disrupt the DNA:RNA hybrid and that required to unfold the stem III pseudoknot: The larger difference gives a low repeat addition processivity while the smaller one gives a high repeat addition processivity.

## 2. Model

In other polymerase enzymes such as DNAP and RNAP, it has been assumed that there exists an ssDNA-binding site in the vicinity of the polymerase site that has a high affinity for the unpaired base (or the sugar-phosphate backbone of the unpaired base) on template DNA [32]. On the analogy of DNAP and RNAP, we make the similar hypothesis for the telomerase. *In TERT, there exists an ssRNA-binding site A adjacent to the polymerase site and the binding site A has a high affinity for the unpaired base (or the sugar-phosphate backbone of the unpaired base) on the template in the RNA component.* This assumption is consistent with the coinmmunoprecipitation assay showing that the template region of *Tetrahymena* telomerase is also important for optimal TERT binding besides other regions such as TBE motif, TRE motif and stem IV loop in the RNA component [20]. As in DNAP [32] and RNAP the strong binding of site *A* to ssDNA can be used to explain the induced-fit mechanism, we can similarly have the following anticipation for the telomerase: The strong interaction of the site *A* with an unpaired base on the template RNA induces the conformational change in residues in the vicinity of the polymerase site that is adjacent to the site *A*. This unpaired-base-related conformational change thus results in the polymerase site having a much higher affinity for the structurally compatible nucleotide than structurally incompatible nucleotides.

It has been structurally determined that in bacteriophage T7 DNAP the nucleotide binding to (releasing from) the polymerase site induces the rotation of the finger domain from open (closed) conformation to closed (open) conformation, with the nucleotide being able to bind to the polymerase site in the open-finger conformation while the closed-finger conformation activating the chemical reaction of nucleotide incorporation [33,34]. In single-subunit phage T7 RNAP the nucleotide binding (releasing) induces the rotation of an $\alpha$ helix, termed the O helix, from open (closed) conformation to closed (open) conformation, with the nucleotide being able to bind to the polymerase site in the open conformation while the closed conformation activating the nucleotide incorporation [35,36]. Similarly, for the telomerase we make the following assumption: *After a nucleotide binds to the polymerase site, the stem IV loop is induced to swing towards the polymerase site from its equilibrium position that is away from the polymerase site, forming the ternary complex of stem IV, nucleotide and polymerase site, which activates the nucleotide incorporation. In the nucleotide-free state, there is no interaction between stem IV loop and the polymerase site and thus the stem IV loop swings away from the polymerase site (back to its equilibrium position), which allows for the nucleotide able to bind to the polymerase site*. The assumption that the swinging of stem IV to the polymerase site activates the nucleotide incorporation is consistent with the previous experiments showing that the telomerase RNP with RNA lacking stem IV exhibited weak nucleotide addition processivity [20–22]. The experimentally determined stimulation of repeat addition processivity by stem IV cooperated with stem III pseudoknot [20–22] will be discussed later.

For the interaction between telomerase and the primer, we also adopt the two-site model as proposed before [26,27], in which the 3' end of the primer is bound on the template region of the RNA via forming the DNA:RNA hybrid while *the 5' end is bound in a separate anchor site that has specificity for telomeric sequence or a G-rich sequence*. Moreover, it is assumed that the residues between the domain or motif containing the anchor site and that containing the ssRNA-binding site *A* behave elastically.

Based on above assumptions and statements, the mechanism for the processive nucleotide and repeat additions by telomerase is described as follows. For convenience of writing, we use *Tetrahymena* telomerase as an example, where the telomerase RNA is a 159-nucleotide transcript and the template region (3'-AACCCCAAC-5') is from nucleotides 51 to 43 [6] (Fig. 1).

**2.1. Processive nucleotide addition**

We begin with the ssRNA-binding site *A* in the equilibrium position binding to nucleotide 48 on the template region of RNA, as shown in Fig. 1(a), where bases 49-51 of template RNA are bound to the 3' end of the primer through Watson-Crick base pairing and the anchor site is bound to the 5' end of the primer. In this equilibrium conformation of the telomerase, stem IV is not far away from the polymerase site and stem III pseudoknot is in the folding conformation. Since either a correct (or matched) dGTP or an incorrect (or mismatched) dTTP can bind to the polymerase site, although a matched dGTP may have a much larger probability to bind, we thus consider the two cases separately.

*2.1.1. The incorporation of a matched base*

The binding of a matched dGTP to the polymerase site induces stem IV loop swinging towards the polymerase site, forming the ternary complex of stem IV, nucleotide and polymerase site [Fig. 1(b)], which activates the chemical reaction of nucleotide incorporation. Upon the completion of nucleotide incorporation and then PPi release, the ssRNA-binding site *A* will bind to the new nearest unpaired base 47 on the template RNA because the previous base 48 where the ssRNA-binding site *A* has just bound is disappeared due to base-pair formation. At the same time the stem IV loop swings away from the polymerase site to its equilibrium position because the polymerase site becomes nucleotide free, as shown in Fig. 1(c).

*2.1.2. The incorporation of a mismatched base*

After a mismatched dTTP instead of dGTP in Fig. 1(a) binds to the polymerase site, the stem IV loop is induced to swing towards the polymerase site [Fig. 1(a')], which activates the chemical reaction of nucleotide incorporation. After the completion of nucleotide incorporation and then PPi release, the polymerase site becomes nucleotide free and thus the stem IV loop swings away from the polymerase site to its equilibrium position, as shown in Fig. 1(b'). In this case, although the sugar-phosphate backbone of the mismatched dTTP was connected to the backbones of the nascent DNA primer, the mismatched base T is not paired with the sterically corresponding base C on the template RNA. Thus the ssRNA-binding site *A* is still binding to the same unpaired base 48. Therefore, there is no movement of the polymerase site relative to template RNA. This is different from the case for the incorporation of a matched base.

In Fig. 1(b'), consider that a dGTP binds to the nucleotide-free polymerase site. This induces stem IV loop swinging towards the polymerase site, activating the chemical reaction of nucleotide incorporation. However, because of the steric obstacle from the previously incorporated sugar-phosphate backbone, the new dGTP cannot be connected to the backbones of the nascent DNA and, thus, the incorporation cannot be proceed, as shown in Fig. 1(c'). Therefore, the nucleotide addition becomes stalled [37], as in other polymerase enzymes such as DNAP [38] and RNAP [39]. The stalling after a misincorporation gives sufficient time for the mismatched base on the 3' end of primer to be cleaved [37].

In Fig. 1(c'), after the mismatched base is excised, the new dGTP becomes able to be incorporated into the sugar-phosphate backbone of the nascent DNA. Then the stem IV loop swings away from the polymerase site and at the same time the ssRNA-binding site *A* binds to the new nearest unpaired base 47 on the template RNA, as shown in Fig. 1(c).

From Fig. 1(c) the next cycle of nucleotide incorporation will proceed until the ssRNA-binding site *A* reaches base 43 on the template [Fig. 1(d)], thus giving the processive nucleotide addition. Note that, as the polymerase site is moved away from its equilibrium position near base 48, the stem IV loop becomes farer and farer away from its equilibrium position when the polymerase is in nucleotide-bound state, which may result in the unfolding of the stem III pseudoknot.

## 2.2. Processive repeat addition

In Fig. 1(d), the binding of a matched dGTP to the polymerase site induces stem IV loop swinging towards the polymerase site, which activates the nucleotide incorporation. After the completion of nucleotide incorporation and then PPi release, the polymerase site becomes nucleotide free and thus the stem IV loop swings away from the polymerase site to its equilibrium position. Because the previous base 43 where the ssRNA-binding site *A* has just bound is disappeared due to base-pair formation, there is no unpaired base left in the template region and the unpaired bases in the TBE region are bound to and buried in the TERT [18]. Thus there is now no interaction between telomerase RNA and the ssRNA-binding site *A*. Moreover, because of the steric restriction of the primer, the motif composed of the binding site

*A* and the polymerase site are not allowed to move in the direction pointed to the 5' end of the primer, as seen in Fig. 1(e). Therefore, there occurs an elastic force, $F_A$, acting on the primer by the motif composed of the binding site *A* and the polymerase site due to the deviation of the motif from its equilibrium position. But this elastic is not large enough to disrupt the DNA:RNA hybrid.

In Fig. 1(e), consider a nucleotide dNTP binding to the polymerase site and the stem IV loop swings towards the polymerase site. This induces the unfolding of the stem III pseudoknot, as shown in Fig. 1(f), thus a force $F_{III}$ occurring to make bases AACC and bases UUGG reform base pairs or to maintain the pseudoknot III in the unfolding state. Now both the force $F_{III}$ and the elastic force $F_A$ are acting on the primer. This is different from the case before the incorporation of the nucleotide paired with the last base 43 on the template, where both $F_A$ and $F_{III}$ occurred after the dNTP binds to the polymerase site are mainly acting on the template rather than the primer because the binding site *A* is bound strongly to the unpaired base 43 on the template. The total force of $F_{III}$ and $F_A$ in Fig. 1(f) may be large enough to disrupt the DNA:RNA hybrid and then the motif composed of the binding site *A* and the polymerase site are returned to the equilibrium position near base 48, to which the binding site *A* will rebind, as shown in Fig. 1(g). Note that, since the anchor site is bound to the 5' end of the primer, during the period from the moment of the unpairing of the DNA:RNA hybrid to the moment of the rebinding of site *A* to the unpaired base 48 and base pairing of bases 49-51 of template RNA with the 3' end of the primer, the primer cannot be detached from the telomerase.

Figure 1(g) is the same as Fig. 1(b) except that the polymerase site is moved relative to the primer by a sequence repeat. From then on the next repeat will proceed, thus giving the repeat addition processivity.

In the above model for the repeat addition processivity, the force $F_{III}$ plays a critical role in the unpairing of DNA:RNA hybrid, a prerequisite for repeat addition processivity. The force $F_{III}$ is resulted from the unfolding of the stem III pseudoknot (or resulted from the potential to make bases AACC and bases UUGG reform base pairs in the unfolding conformation of pseudoknot III) when stem IV is swung towards the polymerase site in the position as shown in Fig. 1(f). That implies that both the stem III pseudoknot and the stem IV are required to generate repeat addition processivity, which is in agreement with the previous experimental results [20].

Moreover, the unfolding of the stem III pseudoknot is induced by the binding of a dNTP to the polymerase site (see Fig. 1(f)). On the one hand, it has been experimentally shown that *Tetrahymena* telomerase has a $K_m$ for dGTP incorporation that is about 10-fold lower than the $K_m$ for incorporation of dTTP [40], which suggests that the polymerase site has a much higher affinity for dGTP than dTTP. In addition, in Fig. 1(e), without the stimulation by the conformational change of residues in the vicinity of the polymerase site due to strong binding of site *A* to an unpaired base on the template, the polymerase site has a much lower affinity for dNTP than that when site *A* is bound strongly to an unpaired base such as in the case of Fig. 1(d) (see the above discussion on induced-fit mechanism). Thus, in Fig. 1(e) the polymerase site will be bound most probable by dGTP with a very low binding rate. On the other hand, for the low dGTP concentration, during the long period before dGTP binding since only the 5' end of the primer is bound to the anchor site the telomerase has a large probability to dissociate from the primer. Therefore, the repeat addition processivity requires much high dGTP concentrations or is stimulated allosterically by dGTP, which is consistent with the previous experimental results [41–43].

## 2.3. Mathematical approach to the processive nucleotide and repeat additions

Based on the above model, the motion of the polymerase site can be described by the following equation

$$\Gamma \frac{dx}{dt} = -\frac{\partial \left( V_A + V_H - V_P - V_{III} \right)}{\partial x} + \xi(t), \qquad (1)$$

where $x$ denotes the position of the polymerase site along template RNA and the viscous load acting on the moving motif composed of the binding site *A* and the polymerase site is contained in the drag coefficient $\Gamma$.

$V_A$ in Eq. (1) represents the interaction potential between ssRNA-binding site *A* and unpaired bases on the template: Before the incorporation of the nucleotide paired with base 48 of the template $V_A$ can be shown in Fig. 2(a) and after the incorporation $V_A$ is changed to that as shown in Fig. 2(b). Then with the processive nucleotide incorporation $V_A$ is changed similarly. After the incorporation of the nucleotide paired with base 44, $V_A$ becomes the one as shown in Fig. 2(c) and then after the incorporation of the nucleotide paired with base 43, $V_A$ becomes the one as shown in Fig. 2(d).

$V_H$ in Eq. (1) represents the interaction potential of the motif composed of the binding site $A$ and the polymerase site with the DNA:RNA hybrid, which is related to the free energy required to disrupt the hybrid. Thus the depth, $E_H$, of the potential $V_H$ can be simply calculated by summing the hydrogen-bonding free energy of each Watson-Crick base pair of the DNA:RNA hybrid. Before the incorporation of the nucleotide paired with base $n$ ($n = 48, 47 \ldots 43$) of the template and after the incorporation of the nucleotide paired with base 43, $V_H$ are shown in Fig. 3, where the free energy of A:T base pair is taken as $E_{AT} = 2k_BT$ and that of C:G base pair is taken as $E_{CG} = 3k_BT$.

$V_P$ in Eq. (1) represents the elastic potential resulted from the deviation of the motif composed of the binding site $A$ and the polymerase site from the equilibrium position, which can be written as $V_P = \frac{1}{2}C_P x^2$, where $C_P$ is the elastic coefficient and $x = 0$ is located in the position of base 48. Due to the relative small movement range of the polymerase site and considering that $C_P$ has a small value, for approximation we take $V_P \approx 0$ here.

$V_{III}$ represents the potential of the interaction to make bases AACC and bases UUGG reform base pairs in the unfolding conformation of pseudoknot III when a nucleotide is bound to the polymerase site. When bases AACC are close to bases UUGG, this interaction potential energy is approximately equal to the free energy required to unfold the stem III pseudoknot. Since in the folding conformation of the stem III pseudoknot the stem IV loop is assumed to position around base 48 of the template, it is considered that when the binding site $A$ binds to base 48 or the nearby bases such as base 47 and base 46 the pesudoknot III is not required to unfold when stem IV swings to the polymerase site, implying that the potential depth of $V_{III}$, i.e., the free energy required to unfold pesudoknot III $E_{III} \approx 0$. However, as the binding site $A$ (or the polymerase site) becomes farer and farer away from base 46, the stem III pseudokonot becomes partially unfolded or completely unfolded when stem IV swings to the polymerase site after a nucleotide bound to it. Thus $E_{III}$ increases gradually from 0 to $E_{III} = 10k_BT$ because in the completely unfolding conformation of the stem III pseudokonot two A:U base pairs (with $E_{AU} = 2k_BT$) and two C:G base pairs (with $E_{CG} = 3k_BT$) are required to unwind. For simplicity, we take $E_{III}$ increasing linearly with the increase of the distance between the binding site $A$ or the polymerase site and base 46, i.e., $E_{III} = 2.5k_BT$ when site $A$ is positioned at base 45, $E_{III} = 5k_BT$ when site $A$ is positioned at base 44, $E_{III} = 7.5k_BT$ when site $A$ is positioned at base 43, $E_{III} = 10k_BT$ when site $A$ or the polymerase site is positioned beyond base 43. Note that when the polymerase site is nucleotide free, $E_{III}$ is always equal to 0 because the stem III pseudoknot is in the folding conformation.

The last term in Eq. (1), $\xi(t)$, is the thermal noise, with $\langle \xi(t) \rangle = 0$ and $\langle \xi(t)\xi(t') \rangle = 2k_BT\Gamma\delta(t-t')$.

## 3. Results

### 3.1. Processive nucleotide addition

From Eq. (1) and the temporal evolution of the potentials $V_A$ (Fig. 2), $V_H$ (Fig. 3) and $V_{III}$ as presented in subsection 2.3, it is explicitly seen that, within one repeat sequence, the polymerase site moves processively along the template RNA with the step size of one base provided that the potential depth $E_A$ is large. One example of the calculated results is shown in Fig. 4. The mean dwell time $t_{dwell}$ between two successive steps can be calculated by using the equation $t_{dwell} = \dfrac{1}{k_b[\text{dNTP}]} + \dfrac{1}{k_c}$. Here $k_b$ is the nucleotide-binding rate, which is dependent on the type of base and nearly independent on the position of base along the template because there is no elastic force acting on the polymerase site. As indicated in the experiment [40], the nucleotide-binding rate $k_b$ for dGTP is much larger than that for dTTP. $k_c$ is the nucleotide-incorporation rate, which is dependent both on the type of base and on the position of base along the template because for different positions the forces $F_{III}$ acting on the polymerase site are different. Generally, for the same type of bases, as the polymerase site moves along template in 3' - 5' direction, $k_c$ decreases due to the increase of $F_{III}$. Therefore, it is predicted that the mean dwell time at positions of bases 45 (A) and 44 (A) should be much larger than the mean dwell time at positions of bases 48 (C), 47 (C) and 46 (C). Moreover, the mean dwell time at position of base 43 (C) should also be larger than the mean dwell time at positions of bases 48 (C), 47 (C) and 46 (C).

It is important to note that, before the incorporation of the nucleotide paired with the last base 43 on the template, because of the presence of large potential depth $E_A$ the backward forces resulted from the potentials $V_A$ and $V_{III}$ together with the thermal noise cannot disrupt the DNA:RNA hybrid (see subsection 3.2). Moreover, because of the large potential depth $E_A$ together with the anchor site bound with the 5' end, the primer has a very small probability to dissociate from the telomerase, thus ensuring a high nucleotide addition processivity.

### 3.2. Processive repeat addition

After the incorporation of the nucleotide paired with the last base 43, since the potential $V_A$ becomes flat, it is noted that the time, $T_d$, taken to disrupt the DNA:RNA hybrid is equal to the first-passage time of the motif composed of the binding site $A$ and the polymerase site reaching the position around bases 48 on the template from position between bases 43 and 42 under the potential of $V = V_H - V_P - V_{III}$. For $V_P \approx 0$, by solving Eq. (1) we approximately have

$$T_d = \frac{1}{f}\left\{\frac{D}{f}\left[\exp\left(\frac{f}{D}d\right)-1\right]-d\right\} , \qquad (2)$$

where $d = 0.34$ nm, $D = k_B T / \Gamma$ and $f = 13 k_B T / (d\Gamma)$ that is obtained from $V_H$ represented by curve $a_7 b_7 c_7 d$ in Fig. 3 and the free energy required to unfold pseudoknot III, $E_{III} = 10 k_B T$. Since $fd/D = 13 >> 1$, Eq. (2) can be approximately simplified to $T_d = \frac{D}{f^2}\exp\left(\frac{f}{D}d\right)$. From Eq. (1) we have $T_d \approx 0.36$ ms for $\Gamma = 5 \times 10^{-11} \mathrm{kg \cdot s^{-1}}$. This time $T_d \approx 0.36$ ms is very short so that the DNA primer has a small probability to dissociate from the telomerase although during this period only the 5' end of the primer is bound to the anchor site, thus ensuring the high repeat addition processivity. For comparison, the time taken to disrupt the DNA:RNA hybrid before the incorporation of the nucleotide paired with the last base 43 is calculated from Eq. (2) as $T_d \approx 1 \times 10^3$ second, where the potential depth of $E_H = 20 k_B T$ (with $V_H$ represented by curve $a_6 b_6 c_6 d$ in Fig. 3), the free energy required to unfold pseudoknot III $E_{III} = 7.5 k_B T$ and binding affinity of site $A$ to the unpaired base 43 is taken as $E_A = 17 k_B T$ (a value that is much smaller than that taken in Fig. 4). The too long time $T_d \approx 1 \times 10^3$ second implies that before the incorporation of the nucleotide paired with the last base on the template the DNA:RNA hybrid is almost impossible to disrupt.

It is interesting to see the effect of extended template on the repeat addition processivity. As in Hardy et al. [41], consider a 7-nt $T_2G_5$ repeat instead of the wild-type 6-nt $T_2G_4$ repeat. For this case, after the incorporation of the nucleotide paired with the last base on the template, the potential depth $E_H$ becomes $26\,k_BT$ and the free energy required to unfold pesudoknot III is still approximately equal to or smaller than $E_{III} = 10k_BT$. Thus, from Eq. (2) we have $T_d \geq 5$ ms, which is more than 14 times of the time $T_d \approx 0.36$ ms for the wild-type template. This implies that the DNA primer has a larger probability to dissociate from the telomerase since during this period only the 5' end of the primer is bound to the anchor site. In other words, the repeat addition processivity is reduced greatly, which is consistent with the experimental result [41]. Similarly, for the repeat having more than 7 bases, the repeat addition processivity will be reduced further. On the contrary, for the case of a 5-nt $T_2G_3$ repeat, after the incorporation of the nucleotide paired with the last base on the template, the potential depth of $E_H$ becomes $20\,k_BT$ and the free energy required to unfold pesudoknot III is $E_{III} = 7.5k_BT$. Thus, from Eq. (2) we have $T_d \approx 0.24$ ms that is close to the value ($T_d \approx 0.36$ ms) in the wild-type case, thus also giving a high repeat addition processivity, which is consistent with the experiment [41].

If we use a mutant template with a 6-nt $T_1G_5$ repeat instead of the wild-type 6-nt $T_2G_4$ repeat, we have the potential depth $E_H = 25\,k_BT$ instead of $E_H = 23\,k_BT$. Then from Eq. (2) and with $E_{III} = 10k_BT$ we obtain $T_d \approx 2$ ms, about 5.6 folds of $T_d \approx 0.36$ ms for the wild-type template, meaning a less repeat addition processivity than the wild-type template. Similarly, if we use a mutant template containing only G base, i.e., with a 6-nt $G_6$ repeat, which gives $E_H = 27\,k_BT$, we obtain $T_d \approx 11.7$ ms, meaning a much reduced repeat addition processivity. The future experiment is expected to test these predictions.

## 4. Discussion

In this work, a model is presented for the nucleotide and repeat addition processivity by the telomerase holoenzyme. In the model, the two subunits telomerase RNA and TERT suffice to give the telomerase activity, which is consistent with the previous experiments showing that the telomerase RNA and TERT provide the minimal subunits required to reconstitute enzymatic activity *in vitro* [44,45]. The stem III pseudoknot and stem IV of telomerase RNA play critical roles in the telomerase activity, where stem IV stimulates nucleotide addition processivity but not repeat addition processivity, while stem III is not essential for nucleotide addition processivity but is essential in conjunction with stem IV for repeat addition processivity. In particular, for a high repeat addition processivity it is required a much higher dGTP concentration than required for processive nucleotide addition within a repeat. All these are consistent with the previous experimental results [20–22, 41–43].

Based on the proposed model, the mathematical approach is developed, by which the dynamic behaviors of *Tetrahymena* telomerase are quantitatively studied. For approximation, the total free energy required to disrupt the DNA:RNA hybrid is simply calculated by summing the free energy of each Watson-Crick base pair of the DNA:RNA hybrid and the free energies of base-stacking interaction between these base pairs, which originates mainly from noncovalent van der Waals interactions between adjacent base pairs [46], have been neglected. By including the effects of base-stacking interaction, which is generally base-sequence dependent [46], the potential $V_H$ should also be sequence dependent. In other words, the repeat addition processivity should be sequence dependent, which means that the variation of the sequence of the template can also affect the repeat addition processivity. Similarly, the base-stacking interaction potentials between the base pairs of the stem III pseudoknot should also be included in the determination of the potential $V_{III}$ that is resulted from the unfolding of the stem III pseudoknot. It is very complicated to calculate precisely the total free energy of the DNA:RNA hybrid and that of the stem III pseudoknot when the effects of base-stacking interaction are included. The precise calculation of these total free energies and their effects on the repeat addition processivity will be the subject of further study.

### 4.1. Non-processive repeat addition

Previous experiments showed that some telomerases such as telomerases isolated from fungi appear to be naturally nonprocessive, generating predominantly short products *in vitro* [47–51]. Based on the present model, this can be easily explained as that the free energy $E_H$ to disrupt the DNA:RNA hybrid is so larger than the free energy $E_{III}$ to unfold the stem III pseudoknot that it takes very long time for the motif composed of the polymerase site to jump over the potential height $E_H - E_{III}$, meaning that the force resulted from potential $E_{III}$ rarely disrupting the DNA:RNA hybrid. In addition, as revealed by analysis using mutant RNAs [52], *Tetrahymena* telomerase appears to be much less processive *in vivo* than *in vitro*. This may be due to that the solution condition *in vivo* is different from that in *vitro* and the solution condition may have significant effects on the total free energy to disrupt the DNA:RNA hybrid and that to unfold the stem III pseudoknot. This thus leads to the potential height $E_H - E_{III}$ *in vivo* much larger than that *in vitro*.

## 4.2. Effects of stem IV on the folding of pseudoknot III and telomerase activity

A hypothesis in the present model is the swinging of the stem IV towards the polymerase site from its equilibrium position that is away from the polymerase site after a nucleotide bound to it. It should be noted that this hypothesis is only applicable to the case that the equilibrium position of stem IV is not very far away from the polymerase site, as in the case as shown in Fig. 1. In contrast, if the position of stem IV is too far away from the polymerase site, as shown in Fig. 5, the short-range interaction between the stem IV and the binary complex of polymerase site and the nucleotide cannot induce the swinging of stem IV towards the binary complex. Similarly, the folding of pseudoknot III via the interaction potential to make bases AACC and bases UUGG form base pairs can only be occurred in the case that the bases AACC are close to the bases UUGG, as shown in Fig. 1*f*, due to the very short-range interactions between the pairing bases. In contrast, if bases AACC are far away from bases UUGG, as shown in Fig. 5, the interactions between the pairing bases are negligible and thus the pseudoknot III cannot be formed. Correspondingly, the stem IV is not appropriately positioned due to the structural restriction, as shown in Fig. 5. In contrast, if the stem IV is positioned correctly (close to the equilibrium position) that is induced by interaction with TERT, as shown in Fig. 1, it may push bases UUGG close to bases AACC, thus resulting in the folding of pseudoknot III via the short-range interactions between the pairing bases. Alternatively, the interaction of stem IV with TERT may induce the conformational change of TERT and thus enhances its interaction with stem I, which makes stem I positioned correctly and in turn induces bases UUGG close to bases AACC. Therefore, the position of stem IV, which is determined by its interaction with TERT, affects both the folding of pseudoknot III and the swinging of stem IV towards the binary complex of polymerase site and the nucleotide, thus affecting the nucleotide incorporation activity but not the nucleotide binding. In other words, the mutant stem IV loop that affects its interaction with TERT affects the position of stem IV, leading to the consequence that the loss of protein-induced base-pairing in the pseudoknot correlated with the loss of telomerase activity, in agreement with the experimental results by Sperger and Cech [22].

## 4.3. Stem IV stimulates telomerase processivity in *trans*

As demonstrated by previous experiments [20,21], *Tetrahymena* telomerase with RNA lacking stem IV (e.g., 1-107 RNA in Lai et al. [20]) exhibited weak nucleotide addition processivity and no repeat addition processivity. However, addition in *trans* of a second telomerase fragment containing stem IV (e.g., 108-159 RNA) greatly enhanced both nucleotide and repeat addition processivity. Based on the present model, this can be understood as follows: Although the 108-159 RNA added in *trans* is not connected to the 1-107 RNA directly, the two RNA fragments are connected together indirectly via the way in which both the fragments bind with TERT (for example, stem I in 1-107 RNA segment binds with TERT [53] and at least the distal loop of stem IV in 108-159 RNA segment binds with TERT [20])[Note 1]. Thus after a nucleotide binds to the polymerase site, the swinging of the stem IV loop towards the polymerase site can still unfold the stem III pseudoknot, ensuring the repeat addition processivity.

## 4.4. Some tests of the model

In order to verify the model, in the future it is expected to test the following argument and predictions: (1) The argument that the nucleotide binding to and releasing from the polymerase site induces the swinging of the stem IV loop towards and away from the polymerase site, respectively, by using the FRET method, where the stem IV loop is labeled with donor (Cy3) and the residue near the polymerase site is labeled with acceptor (Cy5). (2) The prediction that the dwell time at positions of bases 45 (A), 44 (A) and 43 (C) for *Tetrahymena* telomerase are longer than the dwell time at positions of bases 48 (C), 47 (C) and 46 (C). (3) The prediction that, with a mutant template (3'-ACCCCCACC-5') in *Tetrahymena* telomerase RNA instead of the wild-type one, the repeat addition processivity becomes smaller; furthermore, with a mutant template (3'-CCCCCCCCC-5') the repeat addition processivity becomes much smaller. (4) The prediction that the repeat addition processivity is mainly determined by the difference between the free energy required to disrupt the DNA:RNA hybrid and that required to unfold the stem III pseudoknot, with the large difference corresponding to a low repeat addition processivity while the small one corresponding to a high repeat addition processivity.

This work was supported by the National Natural Science Foundation of China.

Note 1. Here TERT is considered to be composed of several domains or motifs: The TRE and the anchor site are located in one domain, stem I and stem IV bind to one domain, while ssRNA-binding site A and the adjacent polymerase site are located in another domain. The connections between domains behave elastically. After binding with telomerase RNA, the TERT and telomerase RNA assemble as part of a holoenzyme capable of telomerase activity.

# Figure legends

Fig. 1. Schematic illustration of the processive nucleotide and repeat additions by the telomerase (see text for the detailed description). For clarity, the telomerase RNA is only drawn schematically and stems I and II are not shown. The ssRNA-binding site $A$ and the polymerase (Pol) site in TERT are denoted by red and blue dots, respectively. The DNA primer is drawn in pink and, for clarity, the mismatched nucleotide or base is drawn in green.

Fig. 2. The interaction potential between ssRNA-binding site $A$ and unpaired bases on the template: Before the incorporation of the nucleotide paired with base 48 on the template, $V_A$ is shown in Fig. 2(a). After the incorporation of the nucleotide paired with base 48 and before the incorporation of the nucleotide paired with base 47, $V_A$ is shown in Fig. 2(b). Then with the processive nucleotide incorporation $V_A$ is changed similarly. After the incorporation of the nucleotide paired with base 44 and before the incorporation of the nucleotide paired with base 43, $V_A$ becomes the one as shown in Fig. 2(c). After the incorporation of the nucleotide paired with the last base 43, $V_A$ becomes the one as shown in Fig. 2(d).

Fig. 3. The interaction potential, $V_H$, of the motif composed of the binding site $A$ and the polymerase site with the DNA:RNA hybrid. Before the incorporation of the nucleotide paired with base 48, $V_H$ is represented by curve $a_1 b_1 c_1 d$. After the incorporation of the nucleotide paired with base 48 and before the incorporation of the nucleotide paired with base 47, $V_H$ is represented by curve $a_2 b_2 c_2 d$. After the incorporation of the nucleotide paired with base 47 and before the incorporation of the nucleotide paired with base 46, $V_H$ is represented by curve $a_3 b_3 c_3 d$. After the incorporation of the nucleotide paired with base 46 and before the incorporation of the nucleotide paired with base 45, $V_H$ is represented by curve $a_4 b_4 c_4 d$. After the incorporation of the nucleotide paired with base 45 and before the incorporation of the nucleotide paired with base 44, $V_H$ is represented by curve $a_5 b_5 c_5 d$. After the incorporation of the nucleotide paired with base 44 and before the incorporation of the nucleotide paired with base 43, $V_H$ is represented by curve $a_6 b_6 c_6 d$. After the incorporation of the nucleotide paired with the last base 43, $V_H$ is represented by curve $a_7 b_7 c_7 d$.

Fig. 4. A typical result for the processive movement of the polymerase site along the template RNA within one repeat sequence with $E_A = 27 k_B T$ and the dwell time between two successive steps $t_{dwell} = 1$ second.

Fig. 5. A schematic illustration of the telomerase RNA with pseudoknot III being not folded and stem IV being far away from the equilibrium position.

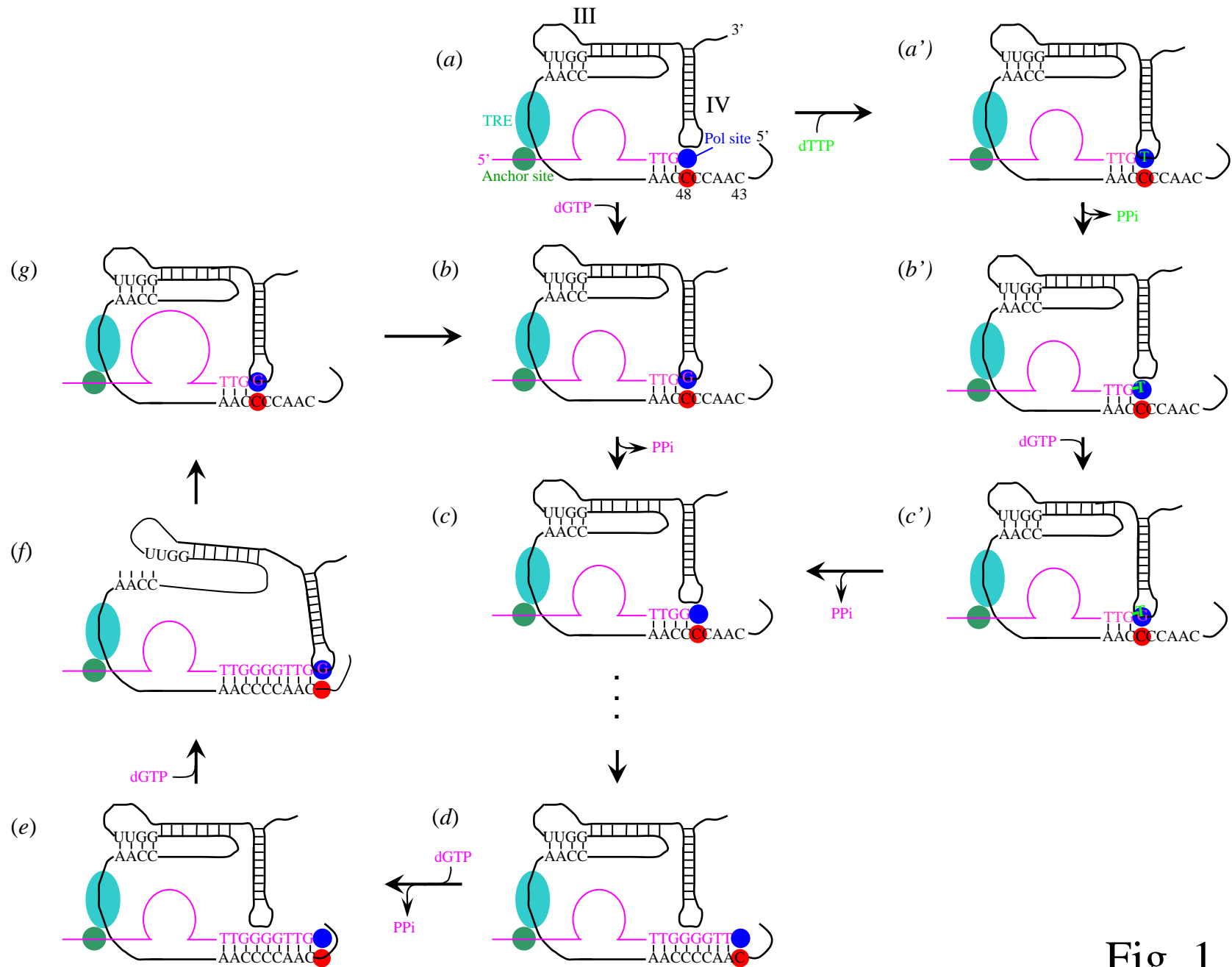

Fig. 1

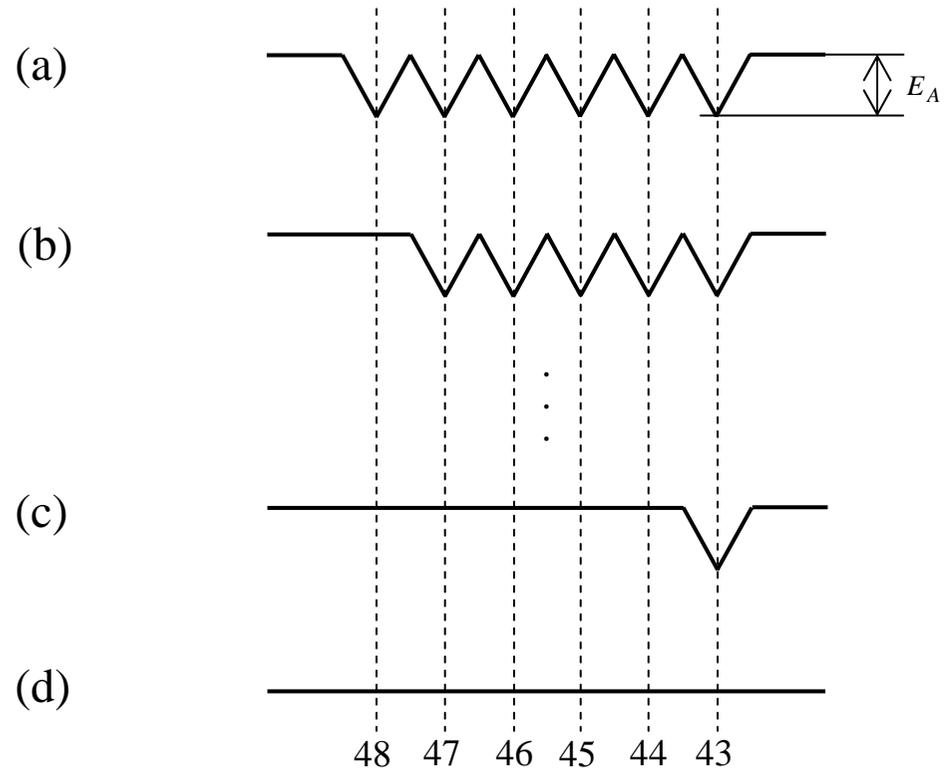

Fig. 2

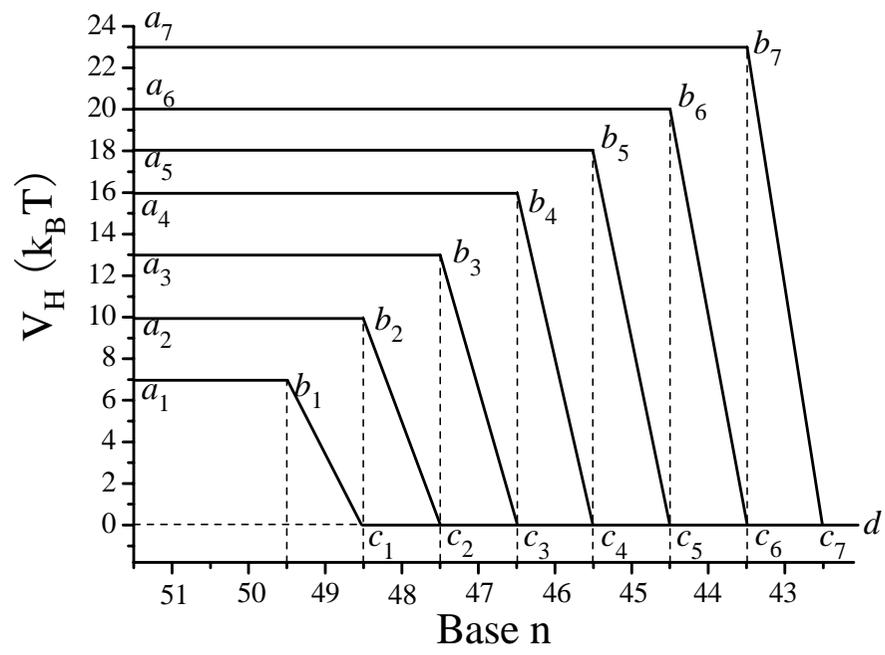

Fig. 3

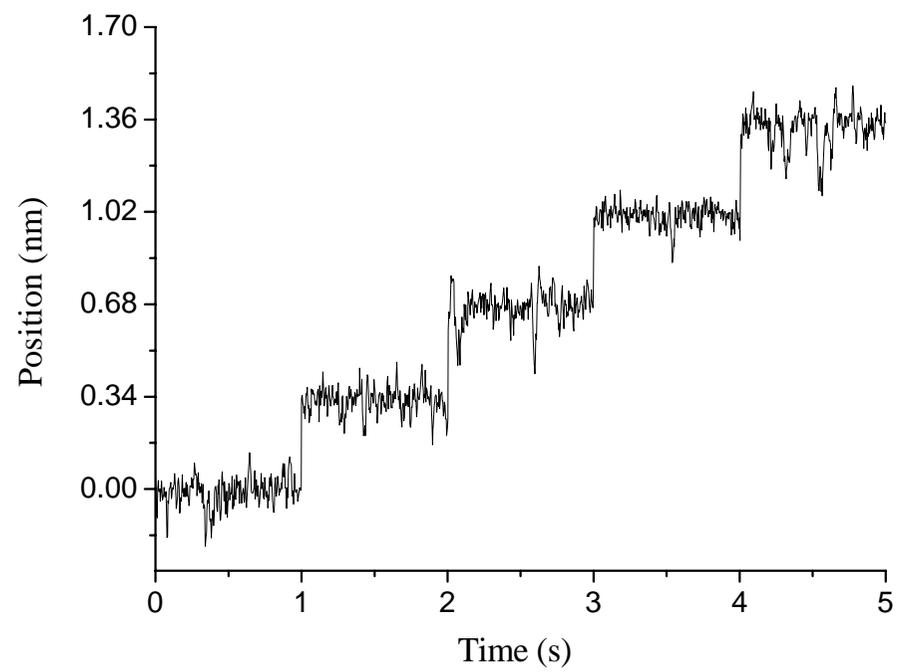

Fig. 4

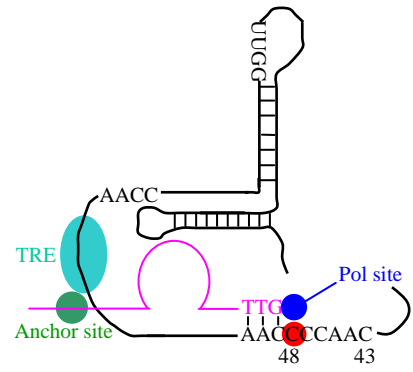

Fig. 5